\documentclass[aip,superscriptaddress,preprint,apl]{revtex4}% Physical Review B
\usepackage{amsmath,amsfonts,amssymb,natbib}
\usepackage[english]{babel}
\usepackage{booktabs,longtable}
\usepackage{graphicx}
\usepackage{dcolumn}
\usepackage{bm}
\usepackage[T1]{fontenc}
\usepackage[latin9]{inputenc}
\usepackage{color}
\usepackage{textcomp}
\usepackage{amsmath}

\begin{document}

\title[]{Van der Waals epitaxy of Mn-doped MoSe$ _{2} $ on mica}

\author{M.T. Dau}
\altaffiliation{These authors contributed equally to this work.}
\affiliation{Univ. Grenoble Alpes, CEA, CNRS, IRIG-SPINTEC, INP Grenoble, 38000 Grenoble, France}

\author{C. Vergnaud}
\altaffiliation{These authors contributed equally to this work.}
\affiliation{Univ. Grenoble Alpes, CEA, CNRS, IRIG-SPINTEC, INP Grenoble, 38000 Grenoble, France}

\author{M. Gay}
\affiliation{Univ. Grenoble Alpes, CEA, LETI, Minatec Campus, 38000 Grenoble, France} 

\author{C. J. Alvarez}
\affiliation{Univ. Grenoble Alpes, CEA, IRIG-MEM, 38000 Grenoble, France} 

\author{A. Marty}
\affiliation{Univ. Grenoble Alpes, CEA, CNRS, IRIG-SPINTEC, INP Grenoble, 38000 Grenoble, France} 

\author{C. Beign\'{e}}
\affiliation{Univ. Grenoble Alpes, CEA, CNRS, IRIG-SPINTEC, INP Grenoble, 38000 Grenoble, France} 

\author{D. Jalabert}
\affiliation{Univ. Grenoble Alpes, CEA, IRIG-MEM, 38000 Grenoble, France} 

\author{J.-F. Jacquot}
\affiliation{Univ. Grenoble Alpes, CEA, IRIG-SYMMES, 38000 Grenoble, France} 

\author{O. Renault}
\affiliation{Univ. Grenoble Alpes, CEA, LETI, Minatec Campus, 38000 Grenoble, France} 

\author{H. Okuno}
\affiliation{Univ. Grenoble Alpes, CEA, IRIG-MEM, 38000 Grenoble, France} 

\author{M. Jamet}
\affiliation{Univ. Grenoble Alpes, CEA, CNRS, IRIG-SPINTEC, INP Grenoble, 38000 Grenoble, France}

\begin{abstract}

  The magnetic order associated with the degree of freedom of spin in two-dimensional (2D) materials is subjected to intense investigation because of its potential application in 2D spintronics and valley-related magnetic phenomena.  We report here a bottom-up strategy using molecular beam epitaxy to grow and dope large-area (cm$ ^{2} $) few-layer MoSe$ _{2} $ with Mn as a magnetic dopant. High-quality Mn-doped MoSe$ _{2} $ layers are obtained for Mn content of less than 5 $\%$ (atomic). When increasing the Mn content above 5$\%$ we observe a clear transition from layer-by-layer to cluster growth. Magnetic measurements involving a transfer process of the cm$ ^{2} $-large doped layers on 100-micron-thick silicon substrate, show plausible proof of high-temperature ferromagnetism of 1 $\%$ and 10 $ \% $ Mn-doped MoSe$ _{2} $. Although we could not point to a correlation between magnetic and electrical properties, we demonstrate that the transfer process described in this report permits to achieve conventional electrical and magnetic measurements on the doped layers transferred on any substrate. Therefore, this study provides a promising route to characterize stable ferromagnetic 2D layers, which is broadening the current start-of-the-art of 2D materials-based applications. 
  
\end{abstract}

%\keywords{2D materials, Mn doping, MoSe$ _{2} $, wet transfer, ferromagnetic order, electrical properties}
\maketitle
%\cleardoublepage

\section{Introduction}
Intentional doping of atomic impurities has been considered as a promising way to induce new functionalities into transition metal dichalcogenides (TMDs)-based two-dimensional (2D) materials. Electron or hole doping via surface adsorption, intercalation or incorporation with electron donors or acceptors, such as NO$  _{x}$, Nb and Li, has been used to tune the Fermi level, to tailor band structures and to engineer structural phase transition of layered TMDs \cite{ Zhao2014, Tedstone2016, Yu2017}. These approaches aimed at making devices based on purely 2D materials such as p-n junctions for highly efficient photodetectors \cite{Choi2014, Sun2018}. Recent efforts to introduce magnetic ordering into 2D materials have been made in the past few years since it will offer room to study the spin physics in 2D materials and potential applications in spintronics and valleytronics \cite{Li2017,Zhao2017,Huang2017,Gong2017,Bonilla2018}. Apart from intrinsically magnetic 2D materials, the spin implantation into TMD materials can be done either by incorporating heteroatoms carrying a magnetic moment or by tuning structural properties via creation of a network of dangling bonds at edge boundaries \cite{Tongay2012, Wang2016, Zhao2017, Ahmed2018, Zhou2018}. As a whole, fabricating magnetic 2D TMD by intentional doping appears to be a straightforward way for further implementation of the 2D TMD into spin and valley-based logic devices and 2D magnetic junctions. Doping would also offer an alternative approach for unravelling the valley-dependent ferromagnetism, i.e. ferrovalleytronics, recently predicted by Tong \textit{et al.} \cite{Tong2016}

Bottom-up growth is one of the techniques to elaborate and dope large-scale TMD layers at the same time. Zhang \textit{et al.} reported stable Mn-doped MoS$  _{2}$ layers by chemical vapor deposition (CVD) and pointed to the role of the substrate on the Mn incorporation. The doping efficiency, the morphology and the quality of the doped monolayers are limited by the reactivity of the substrate surface \cite{Zhang2015}. A high chemical affinity between the dopants and the substrate could promote the dopants to react with the substrate rather than to incorporate into the 2D layers. Another strategy is to tailor the structural and electronic properties in order to induce magnetic phases, which was demonstrated by the intercalation process of Re upon the CVD growth of MoSe$ _{2} $ reported by V. Kochat \textit{et al.} \cite{Kochat2017}. The study showed that the phase transition from 2H to 1T$ ^{'} $ phase as a function of Re doping may occur, giving rise to a significant modification of the electronic structure, hence, to a new magnetic ground state in the doped layers. In the distorted lattice (1T$ ^{'} $ phase) with extra electrons injected from the substitutional Re dopants, the degenerate Mo 4\textit{d} orbitals ($d_{xy} $, $d_{xz} $, $d_{yz} $), which are energetically stable, exhibit total spins higher than that of the trigonal prismatic coordination symmetry (2H phase). Such a 2H-to-1T$ ^{'} $ transition results in correlated magnetism of the Re-doped MoSe$ _{2} $ including ferromagnetic, paramagnetic and antiferromagnetic orderings. 
In this work, we aim to elaborate large-scale Mn-doped few-layers MoSe$ _{2} $ using molecular beam epitaxy (MBE). We developed a growth strategy with co-deposition of Mo and Mn on mica substrate under a Se-rich atmosphere. The opening of Mn flux shutter was carefully controlled upon the deposition of Mo in order to achieve a homogenous doping. The structure of the as-grown layers was characterized by reflection high-energy electron diffraction (RHEED) and Raman spectroscopy. We show that Mn incorporation has no influence on the crystalline structure of MoSe$ _{2} $ when the Mn concentration is less than 5$\%$ (atomic percent), while the surface becomes rough due to a 2D-to-3D growth transition for higher concentrations. The 3D growth is associated with the formation of Mn-rich clusters that exhibit typical RHEED patterns of the $ \alpha $-MnSe phase. The Mn concentration was deduced from the combination of Rutherford backscattering spectroscopy (RBS), energy dispersive X-ray spectroscopy (EDX) and X-ray photoemission spectroscopy (XPS) measurements.  We have also developed a transfer process of the centimeter-scale layers from mica to suitable substrates for further characterizations. Magnetic measurements of monolayers TMD are a challenging task owing to the substrate contribution and a small amount of magnetic matter to measure. The successful transfer allowed us to perform the magnetic measurements using highly sensitive magnetometer superconducting quantum interference device (SQUID). In this study, the magnetic data have shown that Mn dopants induce magnetic states in the doped layers which imply ferromagnetic ordering and paramagnetic states. The paramagnetic feature likely stems from the presence of isolated Mn ions and strongly dominates the magnetic properties at low temperatures. The ferromagnetic states could be clearly distinguished when performing the magnetic measurements at high temperatures. Thereby, we point to a high-temperature ferromagnetism in Mn-doped MoSe$ _{2} $ layers, showing a high interest of the doped 2D layers in 2D spintronics and 2D valley-related physics. Finally, we want to stress that Mn acting as a donor in MoSe$ _{2} $ lattice modifies the Fermi level, thus enhancing the contacting transparency of the 2D layers. Indeed, electrical measurements of the 1$\%$ Mn-doped sample carried out with millimeter and microscopic structures have shown a weak contact resistance with low activation energy compared to that of the undoped layers. Furthermore, the negative magnetoresistance (MR) was observed for both structures with a same magnitude, suggesting that the electrical properties are intrinsic properties of the doped layers at macroscopic or microscopic scales.  Our results demonstrate that magnetic doping of the MoSe$ _{2} $ by use of Mn is a robust strategy to establish the spin degree of freedom in 2D TMD layers, paving new perspective of magnetic 2D materials.

\section{Results and discussion}

Prior to inserting Mn into MoSe$ _{2}$, a systematic study of the growth of undoped 3-monolayer (ML)-thick MoSe$ _{2}$ on mica was conducted at first. All characterizations and results are available in Supplementary Material. We found that good quality of undoped MoSe$ _{2}$ layer could be achieved at the growth temperature of 500 $^{\circ}$C with a Mo flux of 5.2 \AA/min. These conditions were consequently used throughout the Mn doping investigation. 
The crystallinity of the doped 3-ML-thick MoSe$ _{2}$ layers as a function of the dopant content of 1$\%$, 5$\%$ and 10$\%$ was monitored during and after the growth by RHEED shown in Fig. 1a. It is shown that the streaky feature of the diffracted patterns is fairly pronounced in the Mn concentration range less than 5$\%$. For the concentration of 5$\%$ and higher, the RHEED patterns become slightly spotty and the diffracted streaks get broaden, indicating a degradation of the layer quality due to high Mn concentration. A close inspection shows that the spotty appearance related to the surface roughness has a well-defined structure that occurs from the formation of crystalline clusters. The clusters structure resembles to the one of Mn-rich phase, MnSe, which results from the excess of Mn. Indeed, when increasing the Mn content, the spotty RHEED corresponding to the MnSe clusters becomes clearly visible while the streaky features associated with the MoSe$ _{2} $ lattice turns to be significantly faint. The deposition of 100$\%$ of Mn confirms the formation of the MnSe clusters with the diffracted spot mesh corresponding to that of the $ \alpha $-MnSe phase. This phase was reported by D. J. O'Hara \textit{et al.} in growing MnSe$_{x}$ (x = 1, 2) layers on GaSe/GaAs(111) and SnSe$ _{2}$/GaAs(111) substrates \cite{OHara2018}. The RHEED diffraction showing Mn content-dependent transition from 2D to 3D growth implies a margin of optimal Mn concentration for a uniform doping. The concentration margin is determined to be less than 5$\%$, meaning that high-quality Mn-doped MoSe$ _{2} $ layers could be reached in these limits. Over this range, the phase separation induced by a low miscibility of Mn in MoSe$ _{2} $ matrix and chemical affinity of Mn regarding Se give rise to the formation of the $ \alpha $-MnSe phase. Existence of the $ \alpha- $MnSe clusters would be detrimental to the magnetic properties of the doped MoSe$ _{2} $ layer since this phase is known to be not ferromagnetic \cite{Pollard1983}.

Figure 1b shows data obtained by in-plane X-ray diffraction (XRD) of the sample with a Mn concentration of 1$\%$. The $\theta-2\theta$ scans indicate the (h00) and (hh0) peaks of the MoSe$ _{2} $ along 0$^{\circ}$ and 30$^{\circ}$ referring to (0h0) direction of mica. We can index all the diffracted peaks which correspond to the substrate and MoSe$ _{2} $ lattice merely. It means that no other phases are present in the layer. The peak positions of the Mn-doped MoSe$ _{2} $ layer and its mosaicity of about 11$^{\circ}$ (Fig. 1c) are almost identical to that of the undoped layer (Supplementary Material). That agrees well with the RHEED patterns, indicating the MoSe$ _{2} $ lattice remains intact for a low doping concentration.  

	One of the challenges in this study is to determine quantitatively the amount of Mn inserted in the MoSe$ _{2} $ layers. We have performed a series of chemical characterizations in order to address this issue in a proper and convincing manner. We first calibrated the atomic density of Mn in the 3-ML-thick MoSe$ _{2} $ layers using RBS technique. An incident $ ^{4} $He$ ^{+}$ beam is accelerated at an energy of 1.9 MeV and the backscattered helium ions are detected at an angle of 160$^{\circ}$ with respect to the incident beam. The beam current is about 25 nA on the analyzed area. Note that RBS allows quantifying the composition of the layers over millimeter-square regions.  Figure 2a shows typical RBS spectra of the doped layer with a nominal Mn concentration of 10$\%$ and the undoped layer in which we can point out the peaks corresponding to Mn, Mo and Se atoms. The quantitative analysis was then performed using a theoretical simulation with help of the commercial software SIMNRA. For the spectrum of doped layer, the simulation yields the Se/Mo atomic density ratio of 2.16 ($\pm$ 0.1) with a Mn concentration of 12.7$\%$ ($\pm$ 0.4$\%$). In this study, we have carried out the analysis of four different Mn concentrations as a function of nominal Mn amount estimated by quartz microbalance, which permits to do extrapolation for other Mn concentrations.  

	We have shown that the Mn atoms are well present in the doped layers over the analyzed surfaces. We have subsequently employed XPS spectroscopy being a powerful tool for chemical analysis to get further insight into the bonding states of each element in the doped layers, in particular, the oxidation states of Mn.  Figure 2b shows typical high-resolution XPS spectra of the doped layer with a nominal Mn concentration of 10$\%$. The Mo 3\textit{d} core level spectrum shows a major doublet at the binding energies of 228.5 eV and 231.7 eV, referred to MoSe$ _{2} $. The Se 3\textit{d} doublet is found at 54.3 eV and 55.1 eV. These peaks are fully consistent to Mo$ ^{4+} $ and Se$ ^{2-} $ oxidation states of MoSe$ _{2} $ \cite{Abdallah2005}. It is shown that Mo and Se 3\textit{d} peaks both feature shoulders at a higher binding energy. These broad peaks are ascribed to the oxidized states of Mo and Se as observed in the undoped layers \cite{Dau2017}. Note that the oxidization originates from the air exposition of the grain edges, which are mostly Mo terminated. On the other hand, Fig. 2b shows distinct peaks related to the Mn 2\textit{p} core level in the doped layer. The spectrum is composed of a major doublet located at binding energies of 641.3 eV and 653.1 eV and a satellite doublet at 646.9 eV and 658.6 eV, indicating the presence of a ligand effect \cite{Banerjee1998}. The binding energy of Mn peaks shows a shift (+2.4 eV, blue arrow) of oxidation state to higher energy compared to the one of metallic Mn$ ^{0} $. We point to a Mn-Se bonding since we may also ascribe a pair of components of Se 3\textit{d} to this bonding (Fig. 2b) \cite{Wang2006}. Part of the main doublet is probably related to Mn-O bonding, but it is not straightforward to identify the contribution of each item because of the low intensity.

	The fitting yields a Se/Mo ratio of 1.88 $\pm$ 0.28 $\%$ and a Mn concentration of 16.2 $\pm$ 1.2 $\%$. The relative atomic concentrations measured with XPS must be carefully considered due to the depth heterogeneity of the material and the low intensity of the Mn peaks. Furthermore, the EDX analysis of the same sample confirms the presence of Mn over nanometer-square surface as shown in Fig. 2d. Note that the EDX mapping of the Mn element does not show any Mn clustering for the analyzed surface. The in-plane TEM micrograph (Fig. 2c) of the same region shows no clusters or residues. It is also shown that the morphology of the doped layer is comparable to the one of the undoped layer (Supplementary Material). We are unfortunately unable to resolve the location of Mn atoms due to the multilayered character of the doped layers.

	In the following, we describe a transfer process of the as-grown layers based on the hydrophilic character of mica. The process is straightforward and can be easily used for transferring the 2D layers onto any substrate. The process is a crucial step to investigate the magnetic and electrical properties of the doped layers because of the two main reasons. First, as mica (muscovite) contains a fraction of iron atoms that give a significant contribution to the measured magnetic signal. Second, the transfer of the layers onto other substrates permits to perform processes that we are not able to make directly on mica such as e-beam lithography. We employed the varnish as a support layer which plays a role of the transporter of the 2D layers. Similarly to other wet transfer processes, the varnish is first spread out on the doped layers and a sticking solid layer is formed after the solvent is evaporated. The layers + mica substrate are then dipped into deionized water where the water penetrates slowly at the interface between the layer and mica since the dipolar water molecule interacts strongly with the mica top-layer composed mostly of K$ ^{+} $ ions. That enables to lift off the stack (2D layer and varnish) which floats at surface of water. We retrieve and put the stack onto the target substrate by fishing gently. We then bake the transferred layers on a hot plate at 80 $^{\circ}$C for few minutes. The transferred layers exhibit a high homogeneity on target substrate with a small fraction of cracking zones. That was also evidenced by optical microscopy and by mapping Raman peaks which are as intense as the ones of the before-transferred samples.
 
	We performed the magnetic measurements with a SQUID magnetometer. Probing the magnetic state of doped 2D layers is an intriguing issue that requires a great attention in terms of sample handling and data analysis. We were also aware of the high sensitivity of SQUID which makes the data interpretation complicated because of a large substrate volume as compared to the one of the 2D layers. Many magnetic measurements using a SQUID or vibrating-sample magnetometer (VSM) on the 2D TMD layers have been done with thick substrates which presumably exhibit a high susceptibility \cite{OHara2018, Tao2017}, leading to a strong substrate contribution to the final magnetic signal. In order to tackle this issue properly, we first transferred the as-grown layers following the transfer process described above onto Si substrates. They are cut into 5 mm x 5 mm pieces and are sharply thinned down to 100 microns in order to minimize the magnetic contribution from the substrate. Furthermore, for the specimen mounting, we employed a sample holder made of two suprasil rods and a special non-magnetic glue to fix the specimens on the holder. Both the holder and the glue exhibit diamagnetic signal. All precautions for sample handling (such as non-magnetic tweezers) were taken seriously throughout the sample mounting. All the measurements presented here were carried out with magnetic fields applied perpendicular to the sample surface. As shown in Fig. 4a, the raw magnetic signals as a function of the number of transfers exhibit a significant opening of the hysteresis loops when increasing the matter quantity. It clearly turns out that the signal from the substrate is mostly dominated by the diamagnetic component. Such a measurement allows proving the reliability of the transfer process regarding the magnetic characterization and our measured magnetic data.

	Fig. 4b shows the magnetic moment as a function of temperature recorded with $ \mu_{0} $H = 0.5 Tesla. We observe a strong dominance of the paramagnetic component in which the magnetic moment drastically drops when increasing the temperature. For temperatures less than 60 K, the paramagnetic signal is so strong that we are not able to subtract it properly in order to obtain the ferromagnetic component. It is worth noting that a contamination from  paramagnetic impurities apart Mn ions in the samples is also possible at this range of temperature, giving rise to wrong magnetic moment estimations. For instance, the hysteresis curve M-H recorded at 10 K of the 1 $\%$ Mn sample shown in Fig. 4c indicates a magnetic moment as high as 1 x 10$ ^{-5} $ emu up to 4 Tesla. Such a high magnetic moment cannot be used for an accurate extraction of the ferromagnetic component, even with reasonable fitting parameters of the paramagnetic component. On the other hand, for temperatures higher than 60 K, we observe a flat plateau (Fig. 4b) where the paramagnetic component is negligible. Indeed, after subtracting a linear diamagnetic signal, the M-H curve of the 1 $\%$ Mn sample measured at 100 K shows no paramagnetic component (Fig. 4c). We clearly see a \textit{s}-shape hysteresis with a saturation field of around 0.08 Tesla, which indicates a ferromagnetic component solely. In order to evaluate properly the saturated ferromagnetic moment, we measured the field-dependent magnetic moment at 100 K as a function of Mn concentration and the results are displayed in Fig. 4d. We found saturated ferromagnetic moments of about 4 $ \pm \ 1$ $ \mu_{B}$ and  1 $ \pm \ 1$ $\mu_{B}$ per Mn ion at 100 K for the samples with 1 $\%$ and 10 $\%$ of Mn, respectively, by assuming that all doped Mn ions are ferromagnetically coupled. Here, the diamagnetic and residual ferromagnetic signals from the substrate were subtracted for these calculations. The collected magnetic data of the sample with 5 $\%$ of Mn are unfortunately not credible to assess quantitatively the magnetic moment per Mn in this sample owing to an excess of magnetic signal. That might arise not only from the Mn doping but also from some unknown contamination. The \textit{s}-shape feature of the M-H curves of the samples 1 $\%$ and 10 $\%$ Mn were also observed at temperatures higher than 100 K and we anticipate a high-temperature ferromagnetic order in these samples. In spite of the fact that the data of the sample with 5 $\%$ of Mn could not be mined, the magnetic data of the two samples with 1 $\%$ and 10 $\%$ clearly indicate that the  ferromagnetic order exists in the doped layers, demonstrating the successful introduction of spins into the MoSe$_{2}$ layer.

	Theoretical calculations predicted that ferromagnetic ground state would exist in magnetically doped TMD layers in which the spin-spin interaction survives at long range albeit fluctuations due to the dimensionality effect \cite{Mishra2013}. A robust ferromagnetic order might persist up to room temperature with high doping concentration of dopant \cite{Ramasubramaniam2013}. It also turns out that substitutional sites occupied by magnetic dopants such as Mn, Co and Fe are energetically favorable although the incorporation structurally breaks the 2D lattice equilibrium \cite{Dolui2013}. For the substitution arrangement, the origin of ferromagnetism in doped 2D TMD is an open question because the nature of interactions between the localized spins mediated with or without itinerant carriers and the anisotropy effect due to the dimensionality limit remain still unclear.  In diluted magnetic semiconductors, Mn is widely used as a dopant in II-VI, III-V semiconductors and it adopts in general a Mn$ ^{2+} $ charge state which has a high spin (S = 5/2) \cite{Dietl2014}. This state enables long-range interactions through the hybridization of band carriers (holes) and localized spin states (\textit{p-d} Zener model) or/and the indirect exchange coupling (superexchange). By analogy, one could envisage such interactions or a notable overlap of dopant states and the conduction/valence bands of the TMD that enables magnetically long-range coupling in the TMD layers. In this respect, recent experiments pointed out that the Mn doping with Mn$^{2+}$ states resulted in ferromagnetic order in MoS$ _{2} $ nanostructures\cite{Wang2016} and an expected ferromagnetism in doped MoS$  _{2}$ monolayers\cite{Zhang2015}. Advanced theorical calculations anticipated that an antiferromagnetic exchange coupling through the hybridization between Mn 3\textit{d} states and \textit{p} states of Se or S would be the origin of the observation of ferromagnetic alignment of all Mn spins in Mn-doped TMD. Furthermore, the delocalized character of Se 4p states would trigger the FM interactions over long range. As a result, a long-range ferromagnetic order can be expected in Mn-doped MoSe$ _{2} $ with an average magnetic moment of 1.27 $\mu_{B}$ per Mn ion \cite{Mishra2013}. This value is comparable to what we found based on SQUID measurements for the sample with 10 $\%$ Mn. Nevertheless, at this concentration, we clearly demonstrate that a compromise between a good crystalline quality and magnetic properties of the doped layer should be considered for further study. 
 
	In the last section, we focus on the electrical properties of the doped layers on two different substrates: mica substrate with millimeter-sized contacts based on Pt deposited in-situ in the MBE chamber and oxide-on-Si substrate with microscopic Hall bars (Au-based micron-sized contacts) as depicted in Fig. 5a and 5b. These figures also show the four-probe resistivity as a function of temperature for the two structures on mica and silicon substrates (hereafter denoted as structure M and S, respectively). Since the layer thickness \textit{t} $\ll$ probe spacing, we estimated the film resistivity by using \cite{Schroder2006}: $\rho = (\pi/ln2)\times(V/I)\times t$, where \textit{V} and \textit{I} are voltage and current, respectively. We observe that the resistivity increases very slowly when decreasing the temperature down to 50 K. Below 50 K, the resistivity increases sharply. In the low-temperature regime (\textit{T} $<$ 50 K), we estimated the activation energy $ E_{a} \approx$ 7.5 meV and 1.7 meV for the structures M and S, respectively. The activation energies E$ _{a} $ are found to be low for two structures, which is indicative of the same origin of the transport carriers. We suggest that the majority of transport carriers dominating the transport are generated from the Mn donors in the doped layers. These values are comparable to the ones of Mn-doped Ge, GaAs layers, indicating deep level doping of Mn \cite{Woodbury1973, Pinto2005}. At this regime, the carriers get localized, leading to the substantial increase of the resistivity.  In the high-temperature regime, we found that the resistivity of the Mn-doped MoSe$ _{2} $ layer with the structure M is one order of magnitude smaller than that of the undoped layer for the same range of temperature as reported earlier \cite{Dau2017}. That clearly indicates doping characteristic of Mn dopants on electrical properties. 

	Finally, the electrical transport under the application of magnetic fields show identical characteristic with a negative MR at low applied fields for both structures (Fig. 5c, 5d). We do not observe any transition to positive MR as a function of temperature as well as a saturation of MR up to 7 Tesla. These features resemble to the electrical properties of the MBE-grown undoped MoSe$ _{2} $ layers and their origin is still unclear so far \cite{Dau2017}. We can speculate that the Mn impurities should have an impact on the negative MR via isolated magnetic moments carried by Mn ions that would act as spin scattering centers. Therefore, the negative MR could persist up to 200 K without any transition, while a negative-to-positive crossover was evidenced at around 90 K in multilayered flakes TMD \cite{Zhang2016}. A striking feature we want to underline here is that the MR curves of the macroscopic contacts (structure S) have the same magnitude as the ones of the millimeter contacts (structure M) for the same range of temperatures. This finding suggests that the transport properties are intrinsic properties of the doped layers which are independent on the characteristic scale of the measurements. This also bears out the regularity and the feasibility of our transfer process regarding the properties of the doped layers. Unfortunately, we did not observe any magnetic correlation with the electrical data and the Hall measurements show negligible signals for the whole studied range of temperatures. A low mobility of the carriers expected in our layers could be an argument for this absence.

\section{Conclusion}

	To summarize, we have reported the structural, magnetic and electrical properties of the Mn-doped MoSe$ _{2} $ layers grown on mica via van der Waals epitaxy. It has been shown that the crystalline quality of the layers remains unchanged for the Mn concentrations lower than 5 $\%$. However, the Mn dopant has a negative impact on the film quality when increasing the Mn concentration above 5 $\%$, causing the formation of MnSe clusters. A combination of several chemical analysis techniques such as RBS, XPS and EDX has provided robust evidence of the incorporation of Mn atoms as well as its composition in the layers. We have found interesting magnetic and electrical properties in the doped layers: high-temperature ferromagnetic ordering in samples with Mn concentration of 1 $\%$ and 10 $\%$; a low layer resistance as compared to the one of undoped layers. These findings would trigger further investigation of magnetic doping with different dopants and other 2D materials, which is so far a promising alternative route to synthetize stable and high-temperature ferromagnetic 2D materials.

\section*{Supplementary Material}

See Supplementary Material for details of the growth and properties of the undoped and doped MoSe$ _{2} $ layers.

\begin{acknowledgements}
The authors thank Vincent Mareau, Laurent Gonon and Jo\~{a}o Paulo Cosas Fernandes for their assistance and discussions with Raman spectroscopy. The authors acknowledge Yann Almadori and Benjamin Gr\'{e}vin for AFM characterizations; Nicolas Mollard for preparation of TEM samples. This work is supported by Agence Nationale de Recherche within the ANR contracts (MoS2ValleyControl and MAGICVALLEY). The French state funds LANEF (ANR-10-LABX-51-01) and Equipex (ANR-11-EQPX-0010) are also acknowledged for their support with mutualized infrastructure. Minh Tuan Dau acknowledges the support from the Enhanced Eurotalents programm.   

\end{acknowledgements}

%\bibliography{Mica_biblio}

\begin{thebibliography}{}
\expandafter\ifx\csname natexlab\endcsname\relax\def\natexlab#1{#1}\fi
\expandafter\ifx\csname bibnamefont\endcsname\relax
  \def\bibnamefont#1{#1}\fi
\expandafter\ifx\csname bibfnamefont\endcsname\relax
  \def\bibfnamefont#1{#1}\fi
\expandafter\ifx\csname citenamefont\endcsname\relax
  \def\citenamefont#1{#1}\fi
\expandafter\ifx\csname url\endcsname\relax
  \def\url#1{\texttt{#1}}\fi
\expandafter\ifx\csname urlprefix\endcsname\relax\def\urlprefix{URL }\fi
\providecommand{\bibinfo}[2]{#2}
\providecommand{\eprint}[2][]{\url{#2}}

\bibitem{Zhao2014}
P. Zhao \textit{et al.}, ACS Nano \textbf{8}, 10808 (2014).

\bibitem{Tedstone2016}
A. A. Tedstone, D. J. Lewis, and P. O'Brien, Chem. Mater. \textbf{28}, 1965 (2016).

\bibitem{Yu2017}
Y. Yu, G. Li, L. Huang, A. Barrette, Y.-Q. Cai, Y. Yu, K. Gundogdu, Y.-W. Zhang, and L. Cao, ACS Nano \textbf{11}, 9390 (2017).

\bibitem{Choi2014}
M. S. Choi, D. Qu, D. Lee, X. Liu, K. Watanabe, T. Taniguchi, and W. J. Yoo, ACS Nano \textbf{8}, 9332 (2014).

\bibitem{Sun2018}
M. Sun, D. Xie, S. Yilin, L. Weiwei, and R. T. Ren, Nanotechnology \textbf{29}, 015203 (2018).

\bibitem{Li2017}
B. Li, T. Xing, M. Zhong, L. Huang, N. Lei, J. Zhang, J. Li, and Z. Wei, Nat. Commun. \textbf{8}, 1958 (2017).

\bibitem{Zhao2017}
X. Zhao, T. Wang, C. Xia, X. Dai, S. Wei, and L. Yang, J. Alloys Compd.\textbf{698}, 611 (2017).

\bibitem{Huang2017}
B. Huang \textit{et al.}, Nature \textbf{546}, 270 (2017).

\bibitem{Gong2017}
C. Gong \textit{et al.}, Nature \textbf{546}, 265 (2017).

\bibitem{Bonilla2018}
M. Bonilla, S. Kolekar, Y. Ma, H. C. Diaz, V. Kalappattil, R. Das, T. Eggers, H. R. Gutierrez, M.-H. Phan, and M. Batzill, Nat. Nanotechnol. \textbf{13}, 289 (2018).

\bibitem{Tongay2012}
S. Tongay, S. S. Varnoosfaderani, B. R. Appleton, J. Wu, and A. F. Hebard, Appl. Phys. Lett. \textbf{101}, 123105 (2012).

\bibitem{Wang2016}
J. Wang, F. Sun, S. Yang, Y. Li, C. Zhao, M. Xu, Y. Zhang, and H. Zeng, Appl. Phys. Lett. \textbf{109}, 092401 (2016).

\bibitem{Ahmed2018}
S. Ahmed, X. Ding, P. P. Murmu, N.N. Bao, R. Liu, J. Kennedy, J. Ding and J. B. Yi, J. Alloys Compd. \textbf{731}, 25 (2018).

\bibitem{Zhou2018}
Q. Zhou, S. Su, P. Cheng, X. Hu, M. Zeng, X. Gao, Z. Zhang, and J.-M. Liu, Nanoscale \textbf{10}, 11578 (2018).

\bibitem{Tong2016}
W.-Y. Tong, S.-J. Gong, X. Wan, C.-G. Duan, Nat. Commun. \textbf{7}, 13612 (2016).

\bibitem{Zhang2015}
K. Zhang \textit{et al.}, Nano Lett. \textbf{15}, 6586 (2015).

\bibitem{Kochat2017}
V. Kochat \textit{et al.}, Adv. Mater. \textbf{29}, 1703754 (2017).

\bibitem{OHara2018}
D. J. O'Hara, T. Zhu, A. H. Trout, A. S. Ahmed, Y. K. Luo, C. H. Lee, M. R. Brenner, S. Rajan, J. A. Gupta, D. W. McComb, and R. K. Kawakami, Nano Lett. \textbf{18}, 3125 (2018).

\bibitem{Pollard1983}
R. J. Pollard, V. H. McCann, and J. B. Ward, J. Phys. C: Solid State Phys. \textbf{16}, 345 (1983).

\bibitem{Abdallah2005}
W. A. Abdallah and A. E. Nelson, J. Mater. Sci. \textbf{40}, 2679 (2005).

\bibitem{Dau2017}
M. T. Dau \textit{et al.}, Appl. Phys. Lett. \textbf{110}, 011909 (2017).

\bibitem{Banerjee1998}
D. Banerjee and H. W. Nesbitt, American Mineralogist \textbf{83}, 305 (1998).

\bibitem{Wang2006}
L. Wang, L. Chen, T. Luo, K. Bao, and Y. Qian, Solid State Communications \textbf{138}, 72 (2006).

\bibitem{Wu2017}
C.-T. Wu, S.-Y. Hu, K.-K. Tiong, and Y.-C. Lee, Results Phys. \textbf{7}, 4096 (2017).

\bibitem{Tao2017}
L. Tao, F. Meng, S. Zhao, Y. Song, J. Yu, X. Wang, Z. Liu, Y. Wang, B. Li, Y. Wang, and Y. Sui, Nanoscale \textbf{9}, 4898 (2017).

\bibitem{Mishra2013}
R. Mishra, W. Zhou, S. J. Pennycook, S. T. Pantelides, J.-C. Idrobo, Phys. Rev. B \textbf{88}, 144409 (2013).

\bibitem{Ramasubramaniam2013}
A. Ramasubramaniam and D. Naveh, Phys. Rev. B \textbf{87}, 195201 (2013).

\bibitem{Dolui2013}
K. Dolui, I. Rungger, C. Das Pemmaraju, and S. Sanvito, Phys. Rev. B \textbf{88}, 075420 (2013).

\bibitem{Dietl2014}
T. Dietl and H. Ohno, Rev. Mod. Phys. \textbf{86}, 187 (2014).

\bibitem{Schroder2006}
D. Schroder, \textit{Semiconductor material and device characterization}, 3rd ed. (IEEE Press; Wiley, Piscataway NJ; Hoboken N.J., 2006).

\bibitem{Woodbury1973}
D. A. Woodbury and J. S. Blakemore, Phys. Rev. B \textbf{8}, 3803 (1973).

\bibitem{Pinto2005}
N. Pinto, L. Morresi, M. Ficcadenti, R. Murri, F. D'Orazio, F. Lucari, L. Boarino, and G. Amato, Phys. Rev. B \textbf{72}, 165203 (2005).

\bibitem{Zhang2016}
Y. Zhang, H. Ning, Y. Li, Y. Liu, and J. Wang, Appl. Phys. Lett. \textbf{108}, 153114 (2016).


\end{thebibliography}
% Encoding: UTF-8
{}

\cleardoublepage

\section*{Figure legends}

\begin{figure}[hbtp]
\centering
\includegraphics[width=\textwidth]{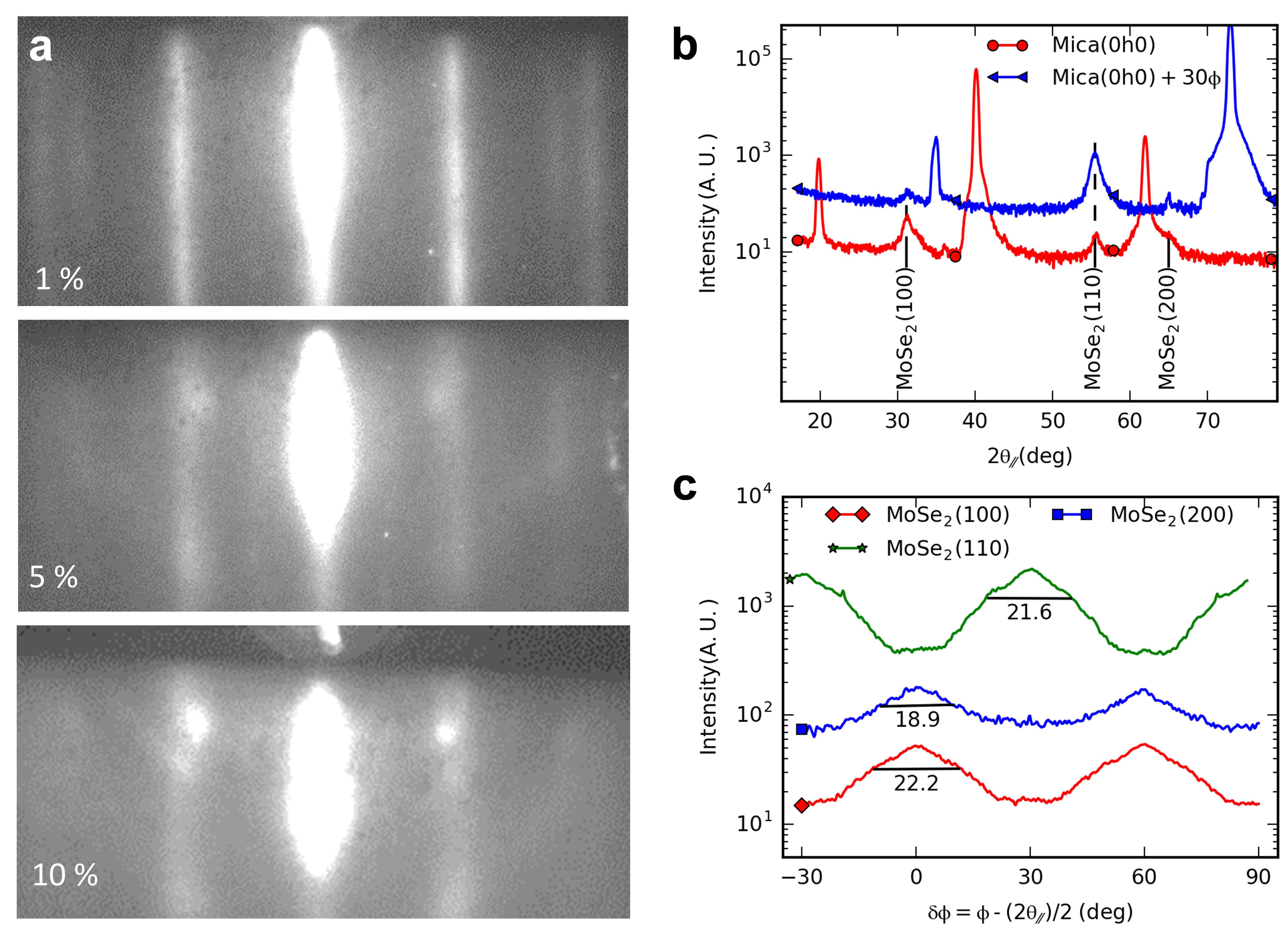}
\caption{(a) RHEED evolution of the doped MoSe$ _{2} $ layers as a function of Mn concentration. (b) XRD $\theta-2\theta$ curves measured at different angles referring to (0h0) direction of mica. (c) $  \Phi$-angle scan along 3 different directions of the MoSe$ _{2} $ lattice.}\label{Fig1}
\end{figure}

\cleardoublepage

\begin{figure}[hbtp]
\centering
\includegraphics[width=\textwidth]{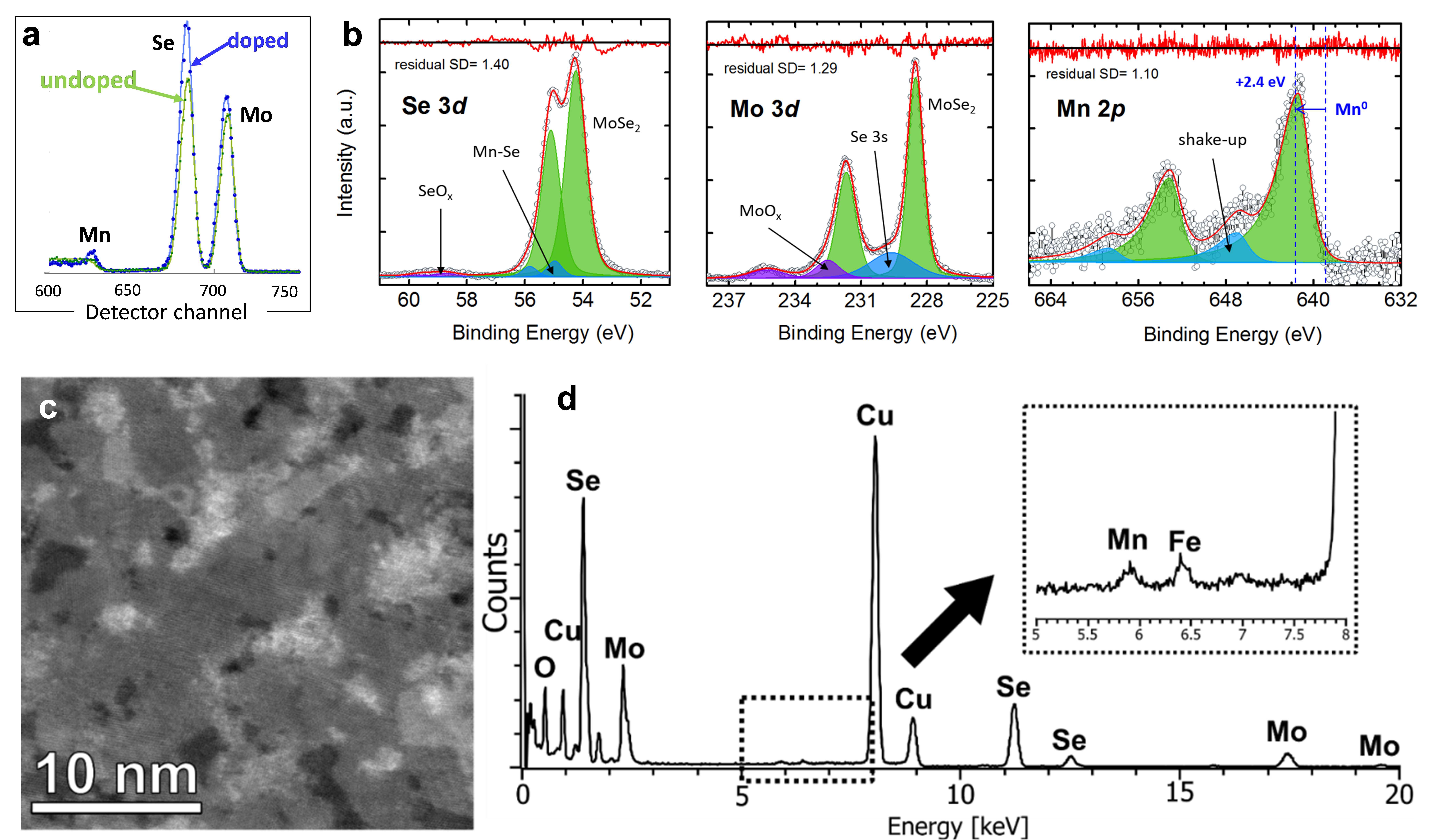}
\caption{Chemical characterization of the doped layers. (a) RBS curves obtained with  undoped (green) and 10$\%$-Mn-doped (blue) samples. (b) Mo 3\textit{d}, Se 3\textit{d} and Mn 2\textit{p} high-resolution XPS core level spectra of 10$\%$-Mn-doped layer. (c) and (d) are EDX data and in-plane TEM micrograph of the 5$\%$-Mn-doped layer at nanometer scale. Note that Cu element detected by EDX comes from the instrument.}\label{Fig2}
\end{figure}

\cleardoublepage

\begin{figure}[hbtp]
\centering
\includegraphics[width=\textwidth]{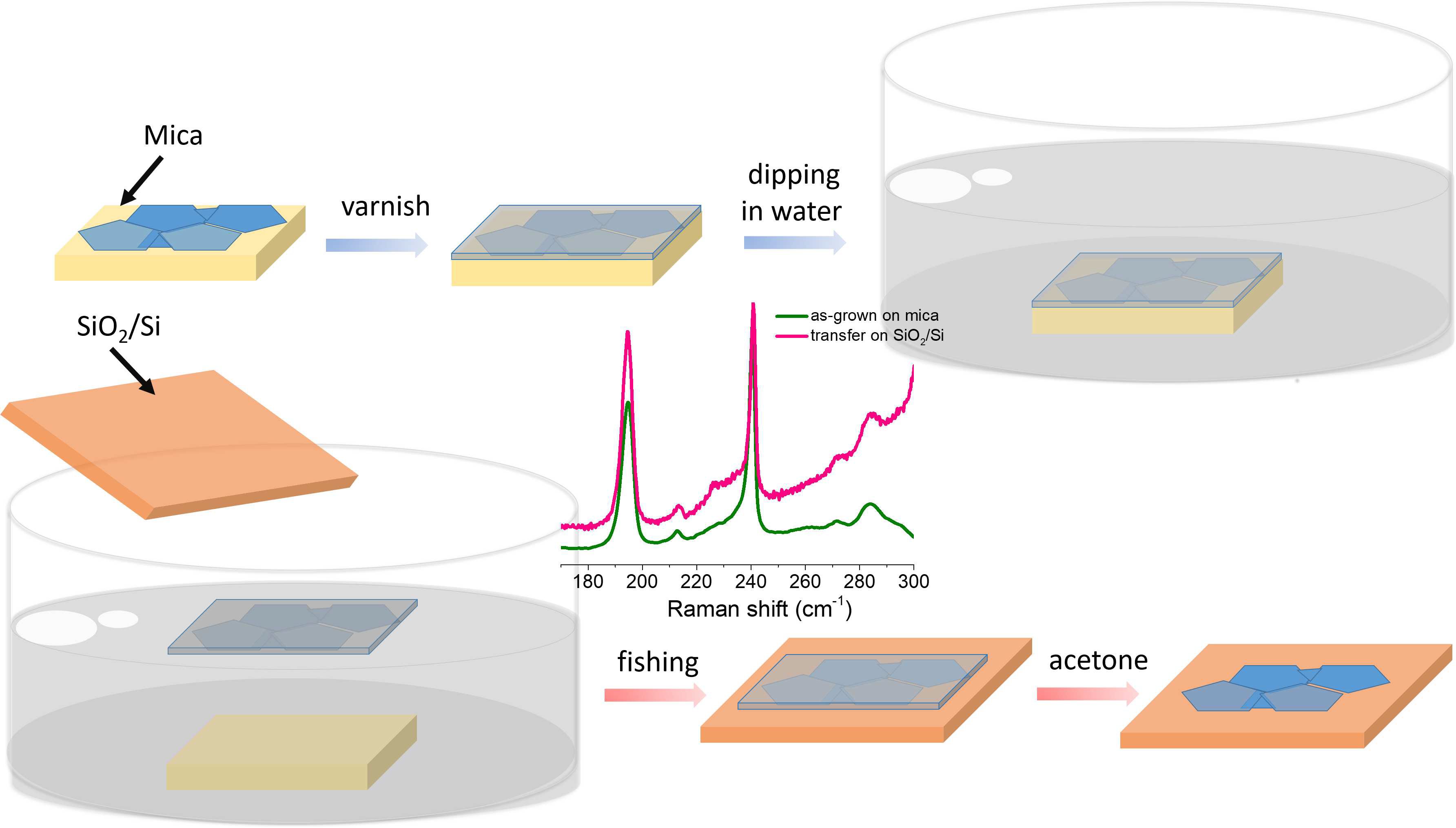}
\caption{Schematics of the transfer process of the as-grown doped MoSe$ _{2} $ layers using varnish and water. The application of a small amount of varnish was done with care to make sure that the surface was covered entirely by the varnish. It takes several minutes in the water until the varnish and MoSe$ _{2} $ layer are removed completely. The removed stack was retrieved with help of targeting substrates. Raman spectrum showing identical peak widths of the as-grown and transferred layers on a silicon oxide substrate indicating comparable quality of the layers.}\label{Fig3}
\end{figure}

\cleardoublepage

\begin{figure}[hbtp]
\centering
\includegraphics[width=\textwidth]{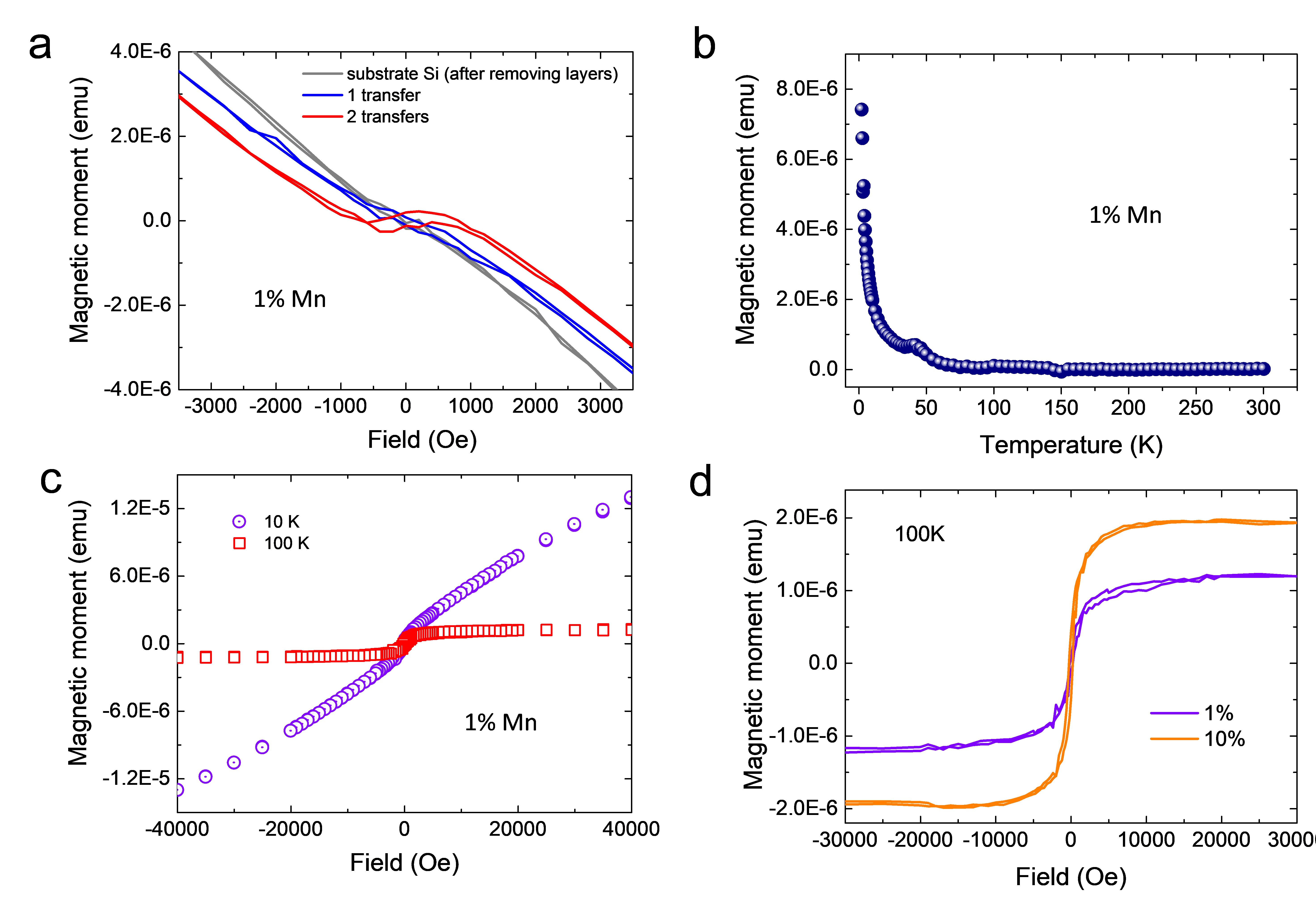}
\caption{SQUID measurements of the doped layers with external magnetic fields perpendicular to the surface normal of all samples. All the data (except curves in (a)) are shown after subtracting the diamagnetic component from the substrate. (a) Raw magnetic field-dependent magnetic moment (M-H) as a function of the number of transfers of the doped sample (1 $\%$). The magnetic intensity is clearly enhanced when we increase the quantity of matter. The substrate was measured after removing all transferred layers using diluted HCl acid. We could confirm that the layers were removed by optical microscopy and Raman spectroscopy. Temperature-dependent magnetic moment (b) and M-H curves recorded at 10 K and 100 K (c) of the doped sample show a strong paramagnetic behavior. (d) M-H curves measured at 100 K of the doped samples with Mn concentrations of 1 $\%$ and 10 $\%$.}\label{Fig4}
\end{figure}

\cleardoublepage

\begin{figure}[hbtp]
\centering
\includegraphics[width=\textwidth]{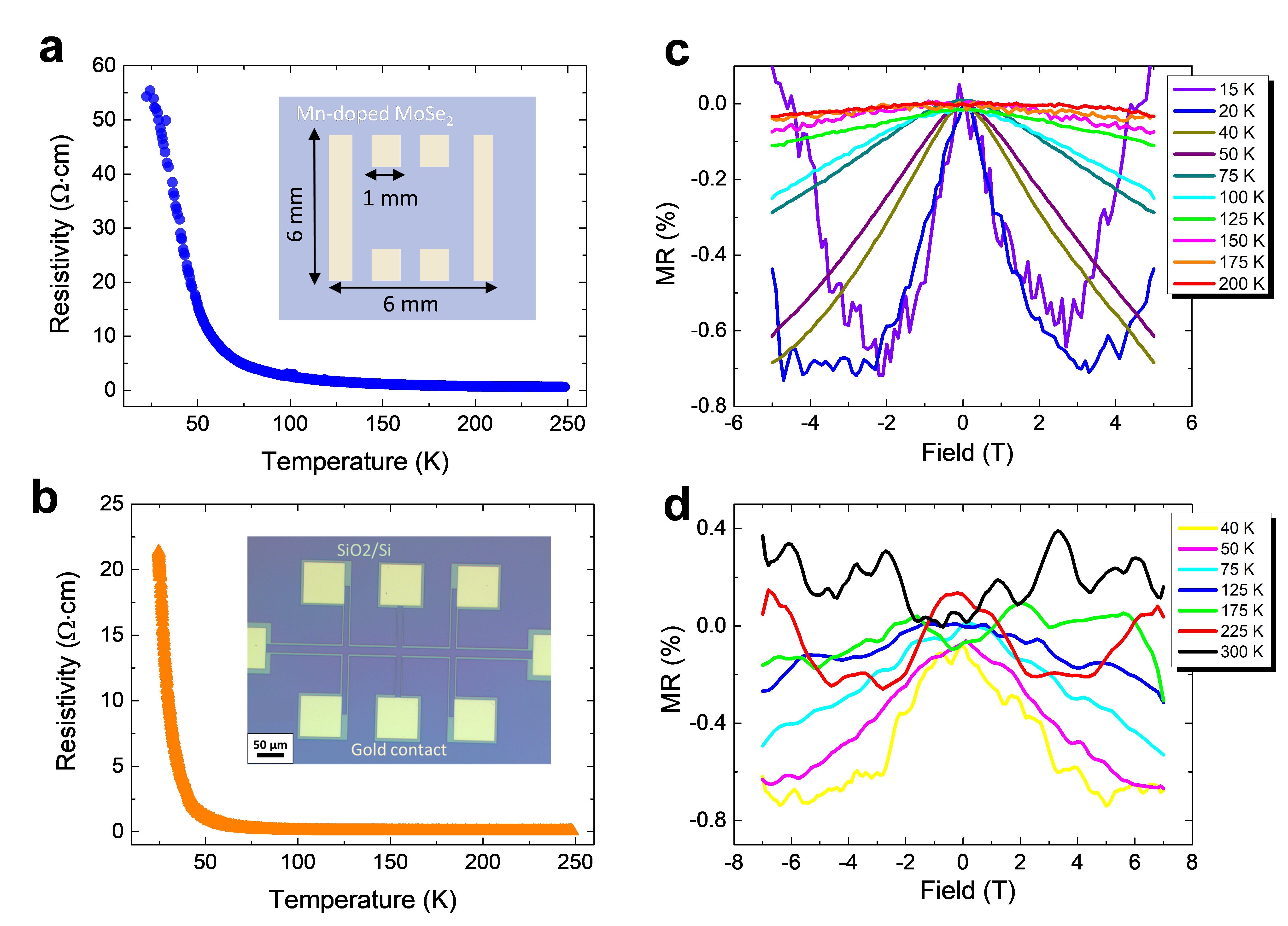}
\caption{Electrical measurements of the doped layer with Mn concentration of 1 $\%$. (a) and (b) are temperature-dependent resistivity with two different structures M and S (see main text) shown in their insets. (c) and (d) show the MR curves at different temperatures obtained with corresponding structures described in (a) and (b), respectively.}\label{Fig5}
\end{figure}

\end{document}